\documentclass[pra,preprint,showpacs]{revtex4}
\usepackage{amsmath}
\usepackage{ulem}
\usepackage{units}
\usepackage{dcolumn}
\usepackage{bm}
\usepackage{graphicx}

\def\be{\begin{equation}}
\def\ee{\end{equation}}
\def\bea{\begin{eqnarray}}
\def\eea{\end{eqnarray}}
\def\etc{{\it etc.}}
\def\Eq#1{Eq.~\ref{#1}}
\def\Eqs#1#2{Eqs.~\ref{#1} and \ref{#2}}
\def\Ref#1{Ref.~\cite{#1}}
\def\Refs#1{Refs.~\cite{#1}}
\def\Refs2#1#2{Refs. \cite{#1} and \cite{#2}}
\def\ie{{\it ie}}

\def\om{\omega}
\def\al{\alpha}
\def\ga{\gamma}
\def\tga{\tilde{\gamma}}
\def\CW{{\cal W}}
\def\CI{{\cal I}}
\def\CN{{\cal N}}

\def\CH{{\cal H}}
\def\CC{{\cal C}}

\def\ket#1{| #1 \rangle}

\def\braket#1#2{\langle #1 | #2 \rangle}
\def\outer#1{| #1 \rangle \langle #1 |}
\def\matrel#1#2#3{\langle #1 | #2 | #3 \rangle}
\begin{document}
	\title{Maximally Entangled States of Four Nonbinary Particles}
\author{Mario Gaeta$^a$, Andrei Klimov$^a$, and Jay Lawrence$^{b,c}$}
\affiliation{(a) Department of Physics, University of Guadalajara, Guadalajara, Jalisco,
Mexico,}
\affiliation{(b) Department of Physics and Astronomy, Dartmouth College, Hanover, NH
03755, USA}
\affiliation{(c) The James Franck Institute, University of Chicago, Chicago, IL 60637, USA}
\date{revised \today}

\begin{abstract}
Systems of four nonbinary particles, each particle having $d \geq 3$ 
internal states, exhibit maximally entangled states that are inaccessible 
to four qubits. This breaks the pattern of two- and three-particle systems, 
in which the existing graph states are equally accessible to binary 
and nonbinary systems alike.  We compare the entanglement properties 
of these special states (called $P$-states) with those of the more familiar 
GHZ and cluster states accessible to qubits.   The comparison includes
familiar entanglement measures, the ``steering'' of states by projective 
measurements, and the probability that two such measurements, chosen 
at random, leave the remaining particles in a Bell state.  These comparisons 
demonstrate not only that $P$-state entanglement is stronger than the other
types, but that it is maximal in a well-defined sense.   We prove that GHZ, 
cluster, and $P$-states represent all possible
%locally equivalent 
entanglement classes of four-particle graph states with prime $d \geq 3$.
%while for composite $d$, additional classes exist with entanglement 
%measures intermediate between $C$ and $P$ states.
\end{abstract}

\maketitle

%\author{Mario Gaeta and Andrei Klimov}

%\author{Jay Lawrence}

\bigskip

\medskip

\section{Introduction}

This work is motivated by the general question of how multipartite entanglement 
develops as both the number of particles and the internal dimension of each 
particle increase.  Our focus  here is on the observation that the four particle 
system represents a sort of transitional case, not only because more than a single
type of graph state \cite{BR1} becomes accessible, but more importantly, because 
the number of such types increases (from two to three) 
on going from the binary to the nonbinary 
cases (all prime $d \geq 3$).  The term graph state is used here to refer to any
nonseparable eigenstate of Pauli operators,  including those of only two or three
particles.  So, for example, two-particle systems exhibit generalized Bell states 
\cite{Durt04,Klimov2,Durt10}, and three-particle systems exhibit generalized 
GHZ states \cite{JL2}, in both cases for any $d$, with these being the only
graph state options.  The situation changes for four-particle systems, first because 
qubit systems can exhibit cluster states ($C$) as well as GHZ states ($G$), and
second, because nonbinary systems (of any prime $d \geq 3$) can access a third
type of graph state, called $P$.  The $P$ states were discovered in the search for 
complete sets of mutually unbiased bases (MUBs) for four qudits \cite{JL2}.  Here 
we document the entanglement properties of the $P$ states in detailed
comparisons with $G$ and $C$ states, and prove that no other types of graph 
states exist for four particles of any prime $d$.

Further motivation is provided by recent ideas for characterizing multipartite
entanglement \cite{Facchi,Arnaud} of pure states through the set of reduced 
density operators associated with all bipartitions; or equivalently, the mixed 
states of all subsystems of up to half the size of the whole system.  If all 
one-particle subsystems are maximally mixed, then the state is called 1-MM 
\cite{Arnaud}, ..., if all $k$-particle subsystems are maximally mixed, 
then it is called $k$-MM.  If a state $\ket{\psi}$ is $k$-MM for $k = [N/2]$ (the 
integer part of $N/2$), then $\ket{\psi}$ is a ``maximally multipartite entangled 
state (MMES)'' \cite{Facchi}.  The following curious situation exists for qubits 
\cite{Arnaud}:  MMES states exist for $N=2, 3, 5$, and 6 qubits, while none 
exists for $N=4$ or any $N \geq 8$ (with the case of $N=7$ unresolved).  A 
central result of this paper is that the $P$ states are MMES for all prime $d$, 
while the $G$ and $C$ states fall short for all $d$; the $P$ states (alone among 
graph states) fill a kind of gap that would otherwise exist for four-particle systems.

We confine our discussion here to graph states because of their prominent place
in practical as well as foundational pursuits, as well as their natural extensions to
nonbinary systems.  There have been extensive studies of entanglement in 
many-qubit systems \cite{Hein}, stimulated in particular by the introduction of 
cluster states and more general graph states \cite{BR1}, with their potential for 
measurement-based quantum computation \cite{BR2,BR3}.  The property of
maximal connectedness, by which graph states can be steered selectively into
entangled final states, was introduced in \Ref{BR1}.  The entanglement 
properties of many-qudit systems ($d\geq 3$) has also received much 
attention \cite{JL2,Durt10,Romero,Wiesniak}, and the advantages of these
nonbinary systems for quantum communication have been pointed out 
\cite{BrussMac, Aravind, DurtQKD} and demonstrated \cite{Grob}.  The stabilizer 
formalism for graph states has been generalized to nonbinary systems
\cite{AshikhKnill,BandB06,Hu}, with useful connections to mutually 
unbiased measurements \cite{Bandyo,Durt10}, which we shall exploit here.

The plan of the paper is as follows:  In the next section we write out the
graph states explicitly and reduce these to their simplest forms, which define
the Schmidt measures.  The reduced forms are used in Sec. III to evaluate  
entanglement measures, and Sec. IV presents an analysis of the steering of 
states by projective measurements utilizing the stabilizer formalism.  While 
the above discussions are self-contained, Appendix A presents a more 
formal approach based on the adjacency matrix, which is then employed in
Appendices B and C to prove that GHZ, cluster, and $P$-states 
%are the only possible locally equivalent classes 
represent all distinct entanglement classes for four-qudit graph states with 
prime $d$.   The concluding Sec. V summarizes the results and discusses
their implications for remaining unanswered questions.

\section{Graph States of Four Nonbinary Particles}

Of all graphs which may be written down for four-qudit systems, only three 
represent 
%locally inequivalent 
distinct entanglement classes, as we prove in Appendices B and C.
The simplest  graphs representative of each class are shown in Fig. 1.Ê 
There is a three-sided alternative for $C$-states, but we declare the 
four-sided graph to be simpler because of its higher symmetry.  
%[*[refto 08 paper by Hu ...?]*]   
\begin{figure}
  \centering
    \includegraphics[scale=.7]{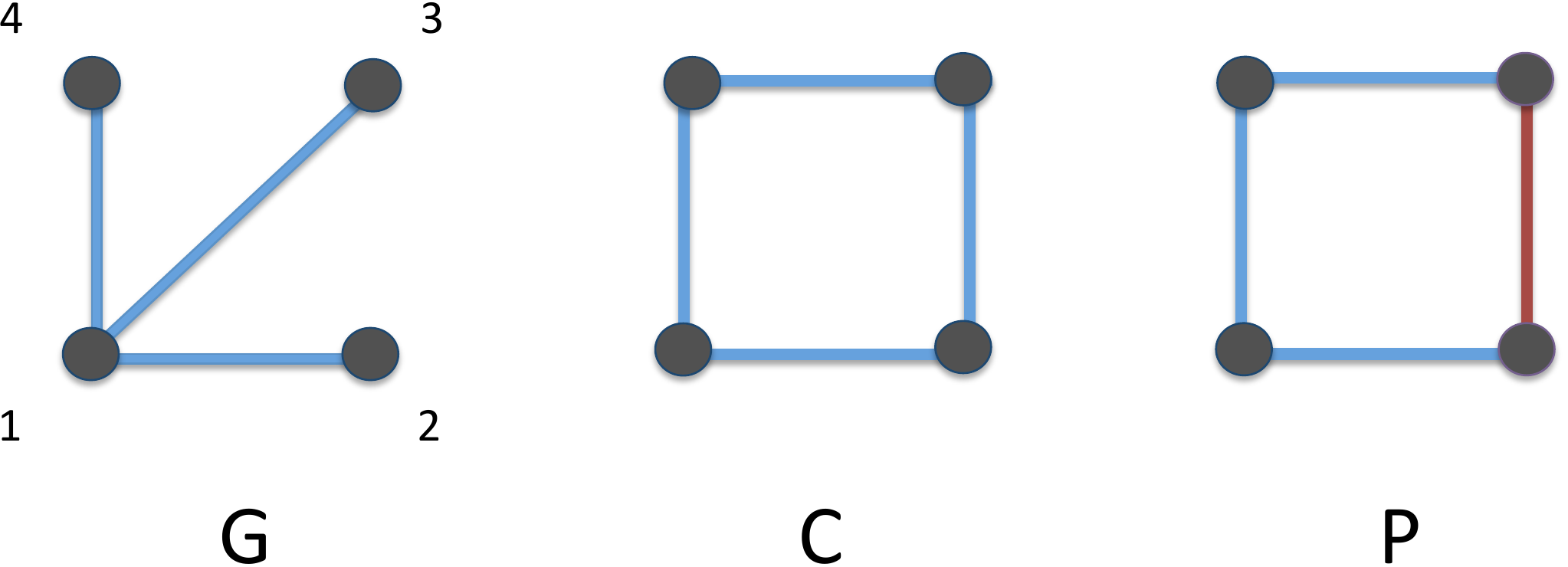}
  \caption{Graphs defining the four-particle GHZ (G), cluster (C), and P 
  states.  Blue edges dictate $\om^{ij}$ factors, red dictate $\om^{-ij}$.}
  \label{Fig1}
\end{figure}

The states are constructed directly from the graphs according to the following 
well-known rules.   One sums over all $d^4$ elements of the standard product 
basis, $\ket{i,j,k,l} \equiv \ket{i}\otimes\ket{j}\otimes\ket{k}\otimes\ket{l}$, 
associating each index with a graph vertex.  Each term in the sum is weighted 
by phase factor dictated by the graph edges.  In our examples a blue edge 
connecting points $i$ and $j$ contributes a phase factor $\om^{ij}$ [with 
$\om \equiv \exp (2 \pi i/d)$], while a red edge contributes $\om^{-ij}$.   
Accordingly, the $G$ state is
\begin{equation}
|G\rangle =d^{-2}\sum_{i,j,k,l}Ê|i,j,k,l\rangle
     \omega^{i(j+k+l)} ,  \label{GHZa}
\end{equation}%
with all four sums extending from 0 to $d-1$.   The $C$ state is
\begin{equation}
|C\rangle =d^{-2}\sum_{i,j,k,l} |i,j,k,l\rangle
     \omega^{j(i+k)+l(k+i)} ,  \label{Ca}
\end{equation}%
and the $P$ state, differing by the color of a single edge, is
\begin{equation}
|P\rangle =d^{-2}\sum_{i,j,k,l}Ê|i,j,k,l\rangle
     \omega^{j(i-k)+l(i+k)} .  \label{Pa}
\end{equation}%
Remarkably, the minus sign strengthens the entanglement for all $d$ except
2, where it simply reproduces the $\ket{C}$ state.

%[[Generalized Bell states have been discussed extensively \cite{Klimov2,
%Durt04}. [[these are Klimov and Durt.; must include ref. 10 prior to this.]]]]  

These expressions can be simplified immediately by recognizing internal
Fourier transforms.  Recall the one-qudit operators,
\begin{equation}
   Z=\sum_{k=0}^{d-1}|k\rangle \omega ^{k}\langle k|   \hskip1.2truecm
    \hbox{and}   \hskip1.2truecm   X=\sum_{k=0}^{d-1}|k+1\rangle \langle k|;  
\label{ZandX}
\end{equation}
these define the one-qudit states $\ket{k}$ as eigenstates of $Z$ with 
eigenvalues $\om^k$, and their Fourier-transforms,
\be
   {1 \over \sqrt{d}} \sum_0^{d-1} \ket{k} \om^{jk} \equiv \ket{j}_x,
\label{Ftrans}
\ee
as eigenstates of $X$ with eigenvalues $\om^{-j}$.   Considering first the
GHZ expression (\ref{GHZa}), one recognizes the sums over $j,k$, and $l$ 
as generating one-qudit states $\ket{i}_x$ for each of qudits 2, 3, and 4.  Now,
locally transforming each of these $X$ states to $Z$ states, one obtains
\begin{equation}
    |G'\rangle = {1 \over \sqrt{d}} \sum_{i}|i,i,i,i\rangle ,  \label{GHZb}
\end{equation}%
where the prime denotes local equivalence to $\ket{G}$ (Eq. \ref{GHZa}).
Similar considerations apply to the $C$ and $P$ expressions, \Eqs{Ca}{Pa},
although now only two of the sums may be interpreted as Fourier transforms, 
not three.  Thus choosing the $j$ and $l$ sums, the second and fourth qudits 
appear in $X$-states with the arguments appearing below.  Locally 
transforming these to $Z$-states, one obtains 
\begin{equation}
    |C'\rangle = {1 \over d} \sum_{i,k}|i,i+k,k,i+k\rangle .  \label{Cb}
\end{equation}%
\begin{equation}
    |P'\rangle = {1 \over d} \sum_{i,k}|i,i-k,k,i+k\rangle .  \label{Pb}
\end{equation}%
It is remarkable that the last two states take the general form
\be
   \ket{\psi(\gamma)} =  {1 \over d} \sum_{i,k}|i,i+ \gamma k,k,i+k\rangle ,
\label{psib}
\ee
and that the states with $\gamma = 0,1$ belong to the same entanglement class 
$\{C\}$, while the $\{P\}$ class contains all others, $\gamma = 2,3,...,d-1$.  
This is proved in Appendices B and C.

The final expressions (\ref{GHZb} - \ref{Pb}) cannot be simplified further, and 
thus make obvious a couple of simple entanglement measures:  The Schmidt 
measure, $M_S$, refers to the minimum number of terms, $N_{min}(\psi)$  in a 
separable basis expansion of $\ket{\psi}$ \cite{Sperling}.  It is defined for 
nonbinary systems by
\be
     M_S = \log_dN_{min}(\psi),
\label{SM}
\ee
and takes the values 1, 2, and 2 for $G$, $C$, and $P$ states, respectively.  
Closely related is the Pauli persistency ($PP$), defined as the minimum
number of one-particle Pauli measurements required to completely remove 
the entanglement \cite{BR1}.   While in general $PP$ is an upper bound on 
$M_S$ \cite{Hein}, it is easy to verify that in our cases (\ref{GHZb} - \ref{Pb}) 
the two are identical.  These simplest measures fail to distinguish between 
the $C$ and $P$ states, but a more quantitative generalized response to 
Pauli measurements (Sec. IV) shows a dramatic difference.  

We should also remark that equations \ref{GHZb} - \ref{Pb} reveal symmetry 
properties not obvious from the graphs themselves:  The GHZ and $P$ 
states are symmetric under the interchange of any two qudits and the 
entanglement is uniformly distributed over all pairs, while the cluster 
states are less symmetric - the 2-4 pair is not equivalent to 1-4 or 3-4.  
It {\it is} equivalent to the 1-3 pair, as one can show by Fourier transforming 
all particles, with the result that 1 and 3 acquire the repeated index in 
place of 2 and 4.   The 
physical consequence of this asymmetry is that particles are more 
entangled with their nearest neighbors on the square graph than with the 
diagonally-coordinated particle.    This concentration of the entanglement 
has been understood to make it more robust than GHZ entanglement 
\cite{BR1}.  The remarkable thing about the $P$ states is that they share 
the uniform entanglement distribution of GHZ states, but with enhanced 
robustness exceeding that of the $C$ states.  These properties will be 
demonstrated quantitatively in the next two sections.

\section{Entanglement}

A thorough yet reasonably concise comparison of entanglement properties is
afforded by the approach of Facchi et. al. \cite{Facchi} and Arnaud and Cerf 
\cite{Arnaud}, which focuses on the reduced density matrices $\rho_A$ and
purities $\pi_A = \hbox{Tr}_A \rho_A^2$ [or equivalently, the linear entropies
$\sim (1-\pi_A)$] of all subsystems ($A$).   We evaluate these quantities for
each of the three states and, at the end, comment on the concurrence and 
wedge product measures, which are both functions of $\pi_A$. 

To begin, consider arbitrary bipartitions 
of pure four-particle states into subsystems $A$ and $B$.  The Hilbert space 
is thus factorized, $\mathcal{H} = \mathcal{H}_{A} \otimes \mathcal{H}_{B}$, 
with subsystem dimensions ($D_A$,$D_ B$) being ($d$,$d^3$) for the 1-3 
partitions and ($d^2,d^2$) for the 2-2 partitions.  A general pure state of the 
system is expanded in the tensor product basis (as in \Ref{Sainz}), 
\begin{equation}
    | \psi \rangle = \displaystyle{\sum_{a=1}^{d_A} \sum_{b=1}^{d_B}}
     c_{a,b}|a \rangle _A \otimes  |b \rangle _B,
\label{psi1}
\end{equation}
where $\{\ket{a}_A\}$ and $\{\ket{b}_B\}$ comprise orthonormal bases in $\CH_A$ 
and $\CH_B$, respectively.   The summation over $b$ at fixed $a$ identifies the
particular state in B, called $\ket{\psi_a}_B$, that is associated with the basis
state $\ket{a}_A$ in $A$.   The set of all such ``associated states,'' both those
in $B$ and their counterparts in $A$, are defined implicitly by the expressions
\be
   \ket{\psi} = \sum_a \ket{a}_A \ket{\psi_a}_B = 
   \sum_b \ket{\psi_b}_A \ket{b}_B,
\label{psi2}
\ee
where obviously $\ket{\psi_a}_B$ and $\ket{\psi_b}_A$ are not normalized.
Defining the reduced density matrices for parts $A$ and $B$ in the usual
manner, $\rho_A = Tr_B \outer{\psi}$ and $\rho_B = Tr_A \outer{\psi}$, and
taking the traces in the bases $\{\ket{b}_B\}$ and $\{\ket{a}_A\}$, respectively,
one obtains
\be
  \matrel{a}{\rho_A}{a'}_A = \braket{\psi_{a'}}{\psi_a}_B \hskip1.3truecm  
  \hbox{and}
  \hskip1.3truecm  \matrel{b}{ \rho_B}{b'}_B = \braket{\psi_{b'}}{\psi_b}_A;
\label{rhome}
\ee
that is, the reduced density matrix of $A$ is identical to the overlap matrix of 
the associated states in $B$, and vice-versa.   Subsystem $A$ is maximally
mixed if and only if 
\be
   \rho_A =  \CI_A/D_A \hskip1truecm \hbox{and hence}  \hskip1truecm
   \pi_A = 1/D_A, \hskip1.5truecm (\hbox{maximal mixing})
\label{MMing}
\ee
where it is understood that $D_A \leq D_B$ (since $\pi_A = \pi_B$, as follows
from \ref{rhome}).

Let us now evaluate $\rho_A$ for all subsystems consisting of one particle 
(we write $A=n$, where $n=1-4$ denotes which particle), and then two particles 
($A = n,m$).   Regarding first the one-particle cases, it is a well-known property of 
graph states that all one-particle subsystems are maximally mixed, so that 
\be
   \rho_n = \CI_n/d  \hskip1.2truecm  \Rightarrow  \hskip1.2truecm
   \hbox{Tr}_n \rho_n^2 = 1/d.
\label{rho1}
\ee
This result is easily confirmed by noting that in all cases, the associated states 
$\{\ket{\psi_a}_B\}$ form orthonormal bases (apart from the factors of $1/\sqrt{d}$) 
in subspaces of $\CH_B$ of dimension $d$.

For the two-particle subsystems ($A = n,m$), we proceed on a case-by-case basis.
We begin with the $P$ states, which (ironically) provide the simplest case.  Choose
$A = 1,3$ for simplicity.  With each basis state $\ket{i,k}$ in $A$, there is a unique 
associated state $\ket{\psi_{i,k}} = \ket{i-k, i+k}/d$ in $B$, so long as $d$ is a prime.  
These associated states form an orthonormal basis (except for the $1/d$ factors) in
$B$, of dimension $d^2$.  According to \Eq{rhome}, $\rho_{n,m}(P)$ is proportional 
to the identity on $\CH_A$, or more explicitly,
\be
  \matrel{ij}{\rho_{n,m}(P)}{i'j'} = \delta_{ii'} \delta_{jj'}/d^2  \hskip1truecm
  \Rightarrow  \hskip1truecm    \hbox{Tr}_{n,m} \rho_{n,m}^2(P) = 1/d^2.
\label{rho2P}
\ee
This result holds for any choice of the pair ($n,m$) and it shows that all such pairs
are in maximally mixed states.   This property makes $P$ states MMES, unique
among four-qudit states.  Let us proceed to see how the other cases fall short.

The $G$ states are less entangled from the current perspective because the 
coefficients $c_{ab}$ in \Eq{psi1} vanish unless the two indices are the same, 
so that the effective basis states in $A$ are $\ket{ii}_{n,m}$, and the associated 
states in $B$ are $\ket{\psi_{ii}}_B = \ket{ii}_{p,q}/\sqrt{d}$, where ($n,m$) and 
($p,q$) label any two distinct pairs.   Clearly the associated states form an 
orthonormal set (except for the $1/\sqrt{d}$ factor), but they span only a 
$d$-dimensional subspace of $B$.  So, although \Eq{rhome} reads concisely as 
$\matrel{a}{\rho_{n,m}(G)}{a'} = \delta_{aa'}/d$, a more explicit reference to the 
full subspace $\CH_A$ reads
\be
 \matrel{ij}{\rho_{n,m}(G)}{i'j'} = \delta_{ii'}\delta_{jj'}\delta_{ij}/d  \hskip1truecm
  \Rightarrow   \hskip1truecm   \hbox{Tr}_{n,m}\rho_{n,m}^2(G) = 1/d.
\label{rho2G}
\ee
This shows that $\rho_{n,m}(G)$ is not maximally mixed because the {\it effective} 
subsystem dimension is less than $D_A$.  Results thus far are collected in Table I.
\begin{table}
\caption{Values of $\pi_A \equiv \hbox{Tr}_A\rho_A^2$ for all one and 
two-particle subsystems ($A$) of four particles in pure $G$, $C$, and $P$ 
states.  Maximal mixing is indicated by minimum values: $1/d$ for one-particle 
subsystems and $1/d^2$ for two-particle subsystems.}
\medskip
%\begin{equation*}
\begin{tabular}{|c|c|c|c|}
\hline
\ State \  & \  $A=n$  \  & \ $n,n+2$ \  & \ $n,n \pm 1$ \  \\ \hline
$G$ & \  $1/d$ \ & \ $1/d$ \ & \ $1/d$ \ \\ 
$C$ & \  $1/d$ \ & \ $1/d$ \ & \ $1/d^2$ \ \\ 
$P$ & \  $1/d$ \ & \ $1/d^2$ \ & \ $1/d^2$ \ \\ \hline
\end{tabular}%
\medskip
%\end{equation*}%
\end{table}

The $C$ states are more complicated because the choice of pairs matters. 
Consider first $A = 1,2$:  With every basis state $\ket{i,i+k}$ in $A$, there is
a unique associated state $\ket{k,i+k}/d$ in $B = 3,4$.   The latter form an 
orthonormal basis of dimension $d^2$, so that, as with $P$ states, 
$\rho_{1,2}(C)$ is proportional to the identity on $\CH_A$.  The identical 
situation clearly arises for another pair, namely $A = 1,4$, so that these two 
pairs are maximally mixed:
\be
   \rho_{1,2}(C) = \rho_{1,4}(C) = \rho_{n,m}(P).
\label{rho2C}
\ee
The remaining pair is $n,m = 1,3$, or equivalently 2,4.  It is simpler to identify the 
latter with $A$, because clearly its effective dimension is only $d$.  So, given the
basis state $\ket{jj}$ in A (where $j \equiv i+k$), the associated state in $B$ is 
$\ket{\psi_{jj}} = \sum_i \ket{i,j-i}_B/d$.  The latter clearly form an orthonormal
set except for the overall constant;  evaluating the orthogonality matrix, one finds
$\matrel{jj}{\rho_{2,4}(C)}{j'j'} = \braket{\psi_{jj}}{\psi_{j'j'}}_B = \delta_{jj'}/d$,
which is identical to the GHZ case leading to \Eq{rho2G}, and so
\be
   \rho_{2,4}(C) = \rho_{n,m}(G).
\label{rho2Cprime}
\ee
Equations \ref{rho2C} and \ref{rho2Cprime} show that while nearest neighbor 
pairs on the square are maximally mixed, diagonally coordinated pairs are not.

In summary, as shown on Table I, all one-particle subsystems ($n$) are 
maximally mixed, so that $G$, $C$, and $P$ states are all 1-MM.  
Two-particle subsystems show a steady progression, but only the $P$
states achieve 2-MM.  Related entanglement measures take their maximum
values in this case.

\subsection{Related entanglement measures}  
We comment briefly on the concurrence and the wedge product measure:
Although these provide different information about mixed states, and may 
evolve differently under dissipative evolution \cite{Sainz}, they reduce to 
functions of the purity when applied to bipartite partitions of pure states.

The concurrence was first introduced \cite{HillW} in the context of pure 
states of two qubits, generalized \cite{Wootters} to mixed states of two qubits, 
and further generalized \cite{Uhlmann,Rungta} to bipartite partitions of 
systems of arbitrary dimension.  It was shown that in the case of pure states,
the concurrence of a bipartition ($A-B$) reduces (within an arbitrary 
multiplicative constant) 
to \cite{Rungta}
\begin{equation}
   \CC = \sqrt{1 - \pi_A},   \label{concurrence}
\end{equation}
noting that $\pi_A = \pi_B$.  So the minimal entries in Table I 
represent maximal values of $\CC$.

The wedge product is a measure of the orthogonality of two associates states; 
its square is defined as
\be
   {\cal W}^2(a,a' ) =  \braket{\psi_a}{\psi_a}_B \braket{\psi_{a'}}{\psi_{a'}}_B
   - |\braket{\psi_a}{\psi_{a'}}_B |^2,
\label{wedge}
\ee
%\begin{equation*}
%\mathcal{W}^{2}\left(a,a' \right) = \det \left[ 
%\begin{array}{cc}
%    \braket{\psi_n}{\psi_n}_B & \braket{\psi_n}{\psi_{n'}}_B    \\ 
%   \braket{\psi_{n'}}{\psi_n}_B &  \braket{\psi_{n'}}{\psi_{n'}}_B%
%\end{array}%
%\right] ;
%\end{equation*}%
that is, $\CW$ is the product of norms times the sine of the angle between the two 
vectors in the subspace of $\CH_B$ that they span.  An entanglement measure 
was first introduced \cite{Meyer} as an average of $\CW^2$ over all one-particle 
subsystems in a many-qubit system, and this was later shown to be equivalent, 
for pure states, to the averaged linear entropies ($1-\pi_A$) of the subsystems 
\cite{Brennen}.  The concept was generalized to arbitrary $k$-particle subsystems 
of many-{\it qudit} systems \cite{Scott}, where, for any particular subsystem ($A$), 
one averages over all distinct pairs of associated states in $B$,
\be
   E_A  =  \displaystyle\sum_{a < a'}{\mathcal{W}^{2}\left( a,a' \right) }.
\label{E}
\end{equation}
One may easily verify that $E_A$ is equivalent to the above measures by 
noting first that the unrestricted sum over all $a$ and $a'$ gives simply $2E_A$, 
because the $\CW(a,a)$ terms vanish.  The double sums are now done 
trivially using Equations \ref{rhome}, demonstrating the identities
\be
   2E_A =  1 - \pi_A = c^2,
\label{identity}
\ee
so that minimal $\pi_A$ values correspond to maximal $E_A$ as well as
maximal $\CC$ values.

\section{Steering by Projective Measurements}

As is well-known, all graph states are maximally connected, in the sense that 
any two particles can be projected into a Bell state by appropriate 
measurements on the others \cite{BR1}. Not every set of measurements 
succeeds, however; some produce product states.  In this section we trace the 
outcomes of all single measurements and all measurement pairs, comparing
the persistency of entanglement as well as the flexibility for producing 
three-particle GHZ states with one measurement, and Bell states with two.
Again we assume that $d$ is a prime, where the existence of complete sets of 
mutually unbiased bases (MUBs) \cite{WF} makes the analysis and the results
more compact.   
 
To introduce the analysis, suppose we prepare the pure four-particle state 
$\ket{\psi^{(4)}}$, and perform a measurement on particle 1 of the observable 
$U$ \cite{observable} (henceforth we will write $U_1$ when we wish to specify
which particle).  If the outcome of the measurement is $\om^i$ ($i = 0,1,...,d-1$), 
then particle 1 is found in the state which we shall call $\ket{U(i)}$, while 
particles 2, 3, and 4 are projected into the associated pure three-particle state
\begin{equation}
| \psi^{(3)} \rangle \sim \langle U_1(i) | \psi^{(4)} \rangle,
\label{project1}
\end{equation}
where ($\sim$) is used because one may wish to regard $\ket{\psi^{(3)}}$ as 
normalized while the right side is not [its norm squared is the probability of the 
measurement outcome ($\om^i$)].  

Expressing these pure states as density matrices, 
$\rho^{(\mu)} = \outer{\psi^{(\mu)}}$, \Eq{project1}  becomes
\begin{equation}
\rho^{(3)} \sim \matrel{U_1(i)}{\rho^{(4)}}{U_1(i)}.  \label{project2}
\end{equation}
The state $\rho^{(4)}$ may be expanded as a sum of its stabilizers, so that
the projected (pure) states, $\rho^{(3)}$ and then $\rho^{(2)}$, are identified by 
their expansions \cite{projection}. The stabilizers form groups which are generated 
by multiplication among any four independent elements, called the generators, 
$g_1, ...,  g_4$.  The stabilizers are thus given by
$S(p_1,p_2,p_3,p_4) = g_1^{p_1}g_2^{p_2}g_3^{p_3}g_4^{p_4}$ \cite{Zhang}, 
where $p_n = 0,1,..,d-1$, \etc, producing a total of $d^4$ stabilizers including the 
identity.  So the stabilizer expansion takes the form of a power series in the 
generators,  
\begin{equation}
    \rho^{(4)} = d^{-4} \sum_{p_1,p_2,p_3,p_4}
(e_1^*g_1)^{p_1} (e_2^*g_2)^{p_2} (e_3^*g_3)^{p_3}(e_4^*g_4)^{p_4},  
\label{powers}
\end{equation}
where the complex quantities, $e_1, ...,  e_4$, are the eigenvalues of the 
corresponding generators, $g_1, ..., g_4$.    This relation is derived in Appendix A 
for interested readers, but one may confirm immediately that $\rho^{(4)}$ is in fact 
an eigenstate of each generator $g_n$ with eigenvalue $e_n$.   

The virtue of this approach is that, if we only care about the type of state and not the 
particular state itself, we need only specify the four generators of $\rho^{(4)}$.  From
these, we may deduce three generators for $\rho^{(3)}$, and from these in turn, two 
generators for $\rho^{(2)}$.  These generator sets alone determine the type of state 
and the nature of its entanglement.

\begin{table}[tbp]
\caption{Generator sets for $G$, $C$ and $P$ states, (a) as taken directly 
from the graphs, and (b) as Fourier transformed with the states $G'$, $C'$, 
and $P'$  of Eqs. \ref{GHZb} - \ref{Pb}.}
\medskip
%\begin{equation*}
\begin{tabular}{|c|ccc|ccc|}
\hline
   & & (a) & & & (b) &     \\
% & \multicolumn{3}{c}{(a) From graphs} & \multicolumn{3}{c}{(b) Fourier transformed}   \\
\hline
   & $G$ & $C$ & $P$ & $G'$  & $C'$ & $P'$     \\
\hline
  $g_1$ & $XZZZ$ & $XZIZ$  & $XZIZ$ &  $XXXX$ & $XXIX$ &  $XXIX$      \\
  $g_2$ & \  $ZXII$ \ & \ $ZXZI$ \ & \ $ZXZ^{-1}I$ \ & 
                  \ $ZZ^{-1}II$ \ & \ $ZZ^{-1}ZI$ \ & \ $ZZ^{-1}Z^{-1}I$ \    \\
  $g_3$ & \ $ZIXI$ \ & \ $IZXZ$ \ &  \ $IZ^{-1}XZ$ \ & 
                  \ $ZIZ^{-1}I$ \ & \ $IXXX$ \ & \ $IX^{-1}XX$ \       \\
  $g_4$ & $ZIIX$ & $ZIZX$ & $ZIZX$ & $ZIIZ^{-1}$ & $ZIZZ^{-1}$ & $ZIZZ^{-1}$     \\
\hline
\end{tabular}
%\end{equation*}
\medskip
\end{table}

Let us briefly describe the generator sets for $\rho^{(4)}$ and then proceed with the 
analysis.  The generator sets inferred directly from the graphs are listed in Table II(a):   
Each generator, $g_n$, identified with its single $X$ factor (implicitly $X_n$), is 
associated with the vertex $n$.  The other factors are powers of $Z$ associated with 
adjacent vertices - those connected by edges.  The power of $Z$ is dictated by the 
edge color:  $Z$ with blue, and $Z^{-1}$ with red.  (These rules are expressed more 
formally in Appendix A.)  Table II(b) shows Fourier transformed generator sets 
compatible with the simplified forms of the graph states (Eqs. \ref{GHZb} - \ref{Pb}). 
These transformations amount to the replacements $Z \rightarrow X$ and 
$X \rightarrow Z^{-1}$, which are applied in the $G$ case to particles 2, 3, and 4; and 
in the $C$  and $P$ cases to particles 2 and 4.  One may easily confirm that the states 
written in Eqs. \ref{GHZb} - \ref{Pb} are indeed joint eigenstates of the four 
corresponding generators, each with eigenvalue unity.

One can see that the full stabilizer sets, as produced by any of the generator sets 
listed in Table II, contain elements in which an arbitrary one-body operator 
($U = X^nZ^m$) appears as the factor associated with any qudit.  That is, writing 
the stabilizers as tensor products, $S = \sigma_1\sigma_2\sigma_3\sigma_4$, 
some stabilizers will have $\sigma_1 = U$, others will have $\sigma_2 = U$, 
and so forth.  If we measure $U_1$, for example, then all those stabilizers having 
$\sigma_1 = U$ (or a power $U^k $), will survive the projection in \Eq{project2}, with 
coefficients ($\om^{ik}$) depending on the measured value ($\omega^i$).   All other 
stabilizers will be annihilated by the projection, because the expectation value of any 
one-body operator ($V$) in an eigenstate of another ($U$, {\it  not} a power of $V$) 
vanishes when the corresponding one-particle bases are mutually unbiased 
\cite{Bandyo, WF}:
\be
   \matrel{U(i)}{V}{U(i)} = \sum_{j=0}^{d-1} \braket{U(i)}{V(j)} \om^j 
   \braket{V(j)}{U(i)} = {1 \over d} \sum_{j=0}^{d-1} \om^j = 0.
\label{expvalue}
\ee

In the following examples, we shall consider measurements by all distinct one-body 
operators (excluding their powers, which would be redundant).   Thus, the measured 
operator $U$ will be $X$, $Z$, or ($XZ^k$), where $k = 1,2,...,d-1$; a total of $d+1$ 
choices for each particle.  With each input state to follow, we consider the outcomes
of all such first measurements, and then (conditionally) of all possible second 
measurements.

\subsection{GHZ states}

Suppose that $\rho^{(4)}$ is any GHZ state expanded in the generators of Table II(b);
the choice of eigenvalues is immaterial.  Consider $X$ measurements.  Because 
$G$ states are symmetric under particle permutations, the results will not depend 
on the choice of particle measured, and so for convenience we choose $X_4$.  To 
determine which stabilizers survive the projection of \Eq{project2} 
and contribute to $\rho^{(3)}$, it suffices to look at just the generators:  Clearly the
first three survive (because powers of $X$ include the identity), leaving $XXX$,  
$ZZ^{-1}I$, and $ZIZ^{-1}$.  These three form a commuting generator set for the 
stabilizers that make up $\rho^{(3)}$, and we recognize it as a GHZ set \cite{GHZ3}.  
Thus, we know that $\rho^{(3)}$ is a $G$ state without carrying the analysis further.   
As a check, measuring $X$ on another particle ($m$) produces the same outcome, 
because products among $g_2$, $g_3$, and $g_4$ produce identity factors $I_m$ 
on the same particle.  Next consider $Z$ measurements:  If we measure $Z_1$, the 
projection (\ref{project2}) produces inverses of $ZII$, $IZI$, and $IIZ$, which qualify 
as generators for $\rho^{(3)}$. These three indicate a product state and the 
removal of all entanglement.  Clearly the same outcome results from a $Z$ 
measurement on any other particle.  Finally, consider the remaining cases, 
$U = (XZ^k)$ ($k \neq 0$).   We can generate stabilizers that associate the factor U 
with any particle we like; for example, $g_1g_2^{-k}$ associates it with the fourth.  
A measurement of $U_4$ then projects from this $(XZ^{-k})XX$, in addition to 
$ZZ^{-1}I$, and $ZIZ^{-1}$, which form a generator set for $\rho^{(3)}$.  Clearly 
the measurement of $U$ on any other particle ($m$) projects a similar 
set of operators, one containing ($X_mZ_m^{-k}$) as a factor among $X$s, the 
others containing factors of $Z$, $Z^{-1}$, and $I$.  Clearly these comprise GHZ
generator sets.  The upshot is that all one-particle measurements, with the 
exception of $Z$ measurements, produce three-particle GHZ states; the $Z$ 
measurements produce product states.   To put this in representation-free terms, 
GHZ entanglement is vulnerable to one out of the $d+1$ possible measurement 
bases.

This vulnerability remains in place for the second measurement, because the
operators $ZZ^{-1}I$, and $ZIZ^{-1}$ are always present in the expansion of 
$\rho^{(3)}$, whatever the first measurement was.  As a result, a second 
measurement of $Z_n$ on any qudit $n$ will project (among others) the 
operators $ZI$ and $IZ$, which qualify as generators for $\rho^{(2)}$ and 
identify a product state.  All other measurement choices, $X$ or $V=(XZ^l$)
~$(l = 1,...,d-1)$, will project $ZZ^{-1}$ and a second operator of the
form $(XZ^{-i})(XZ^{-j})$, where $i$ and $j$ may take the values 0, $k$, or $l$, 
depending on what the first measurement was.  These two generators produce 
(generalized) Bell states \cite{Klimov2,Durt04}, the basic example being
\be
   \ket{\psi} =  {1 \over \sqrt{d}} \sum_i \ket{i,i}.
\label{Bell}
\ee
One may verify that indeed this is an eigenstate of the two generators above, 
whatever specific value the second generator takes (if the second generator 
is $XX$, for example, then the eigenvalue is unity).

Putting together the above outcomes:  Out of all first measurements, a fraction
$d/(d+1$) project three-particle GHZ states.  Out of all second measurements 
on these, the same fraction $d/(d+1)$ project Bell states, their two-particle 
analogs.  So $d^2/(d+1)^2$ of all combined measurements produce Bell states.  
The remaining fraction, $(2d+1)/(d+1)^2$, produce product states.  These 
outcomes are summarized on Table III.  
\begin{table}
\caption{(a) Distribution of first-measurement outcomes resulting from the $d+1$ 
distinct measurements on each particle, reflecting single vulnerable bases for 
the $G$ and $C$ states, and (b) distribution of two-measurement outcomes 
resulting from $(d+1)$ second measurements on each remaining particle 
[total number of measurement sequences is 12($d+1)^2$].}
  \begin{equation*} 
    \begin{tabular}{|c||c|c|c|}
\multispan{4 \hskip0.6truecm \hbox{(a) first measurements}} 
\smallskip \\  
\hline
 \  input  \  & \  $\pi^{(3)}$ \   & \ $S_nB$ \  & \ $G^{(3)}$ \  \\ \hline
$G$ & $4$ & $0$ & $4d$ \\ 
$C$ & $0$ & $4$ & $4d$ \\ 
$P$ & $0$ & $0$ & \ $4(d+1)$ \ \\ \hline
\end{tabular}
\end{equation*}
\begin{equation*} 
    \begin{tabular}{|c||c|c|}
\multispan{3  \hskip.8truecm  \hbox{(b) measurement pairs}}  
\smallskip \\  
\hline
 \ input \ & \  $\pi^{(2)}$ \   & \ \ $B$ \ \  \\ \hline
$G$ & $24d+12$ & $12d^2$ \\
$C$ & $20d+8$ & $ \ 12d^2+4d+4$ \ \\ 
$P$ & \ $12d+12$ \ & $12d^2+12d $ \\  \hline
\end{tabular}%
\medskip
\end{equation*}
\end{table}

\subsection{P states}

The analysis for $P$ states is the simplest because, although it is not immediately
obvious from the generators, there is no stabilizer with more than a single $I$
factor.  This point was made in \Ref{JL2}, but we give a concise proof here in 
Appendix C.  As a consequence, the state $\rho^{(3)}$ resulting from any Pauli
measurement, on any qudit, consists of stabilizers with the same property, and 
is therefore nonseparable \cite{nonseparable}.  The only nonseparable 3-particle 
joint eigenstates of Pauli operators are GHZ states \cite{JL2}.  Therefore, 
remarkably, every first measurement choice projects a 3-particle GHZ state.   

Since all second measurements must be on GHZ states, then, according to 
the arguments given above, $d/(d+1)$ of these project Bell states, while the 
remaining fraction, $1/(d+1)$ project product states.  These fractions are
reflected in Table III(b). 

Remarkably, however, there is no a priori basis of vulnerability for $P$ states.  
Rather, the basis vulnerable to second measurements is determined by the 
basis chosen for the first measurement, and also the choice of qudit.  These 
conditional vulnerabilities arise from the variety of GHZ states projected by 
first measurements:  Every basis appears as the vulnerable one as the 
result of some first measurement.

\subsection{Cluster states}

Cluster entanglement has less a priori vulnerability to measurement bases 
than GHZ entanglement, but such vulnerability exists and it is more complicated.  
We examine the special cases where this enters by considering two revealing 
stabilizers, each of which contains two $I$ factors, namely 
$g_2g_4^{-1} = IZ^{-1}IZ$ and $g_1g_3^{-1} = XIX^{-1}I$.  These identify $Z$ 
as the vulnerable basis for particles 2 and 4, while $X$ is vulnerable for 1 and 3.  
Specifically, a measurement of $Z_2$ projects $IIZ$ and $XX^{-1}I$ from the 
two stabilizers above, and $ZZI$ from $g_2$ itself.  This means 
that particles 1 and 3 are projected into a Bell state, leaving particle 4 in a pure 
eigenstate of $Z$.   This type of state is labelled $S_4B$ \cite{JL2}, identifying the 
separated particle.  By similar arguments, a measurement of $Z_4$ produces a 
state of type $S_2B$. The corresponding vulnerable basis for particles 1 and 3 is 
$X$:  A measurement of $X_3$ projects $XII$ and $IZ^{-1}Z$ from the above 
stabilizers, and $IXX$ from $g_3$ itself.  Thus, the projected state is of the 
type $S_1B$.  A measurement of $X_1$ projects a state of type $S_3B$.   
In all cases, the separating particle is diagonally opposed to the measured 
particle on the square graph.  These are the four special cases of enhanced
vulnerability to second measurements.  It is easy to show that all other first 
measurements, comprising a fraction $d/(d+1)$ of the total number, produce 
3-particle GHZ states, as reflected on Table III.

When the first measurement projects a GHZ state, then, as in the previous cases,
$d/(d+1)$ of the second measurements produce Bell states, while the remaining
$1/(d+1)$ produce product states.  In the four remaining special cases involving 
$S_nB$ states, a measurement on either of the Bell state particles removes the 
entanglement, while a measurement on the separated particle preserves the 
Bell state of the other two.  Comparing these vulnerable cases with those of
GHZ states, rather than losing all entanglement to first measurements, we only 
make it more susceptible to second measurements, where it is lost in 2/3 of the 
cases, but preserved in the remaining 1/3.  As a result, the fraction of successes 
is increased by $1/3(d+1)$ over that of the GHZ case, as reflected in Table III.

To illustrate how the combination of all measurement pathways leads to the
final results, Fig. 2 shows the trajectories taken by 48 distinct measurement
pairs in the special case of $d=3$.  There are actually $12(d+1)^2 = 192$ 
distinct ordered pairs, but the outcome statistics are independent of 
which qutrit is chosen for the first measurement, so the number is effectively 
reduced to 48.  The outcomes after two measurements show a steady 
increase in robustness of entanglement as we progress from $G$ to $C$
to $P$, as well as in steering flexibility for producing arbitrary Bell states.
\begin{figure}
  \centering
    \includegraphics[scale=.7]{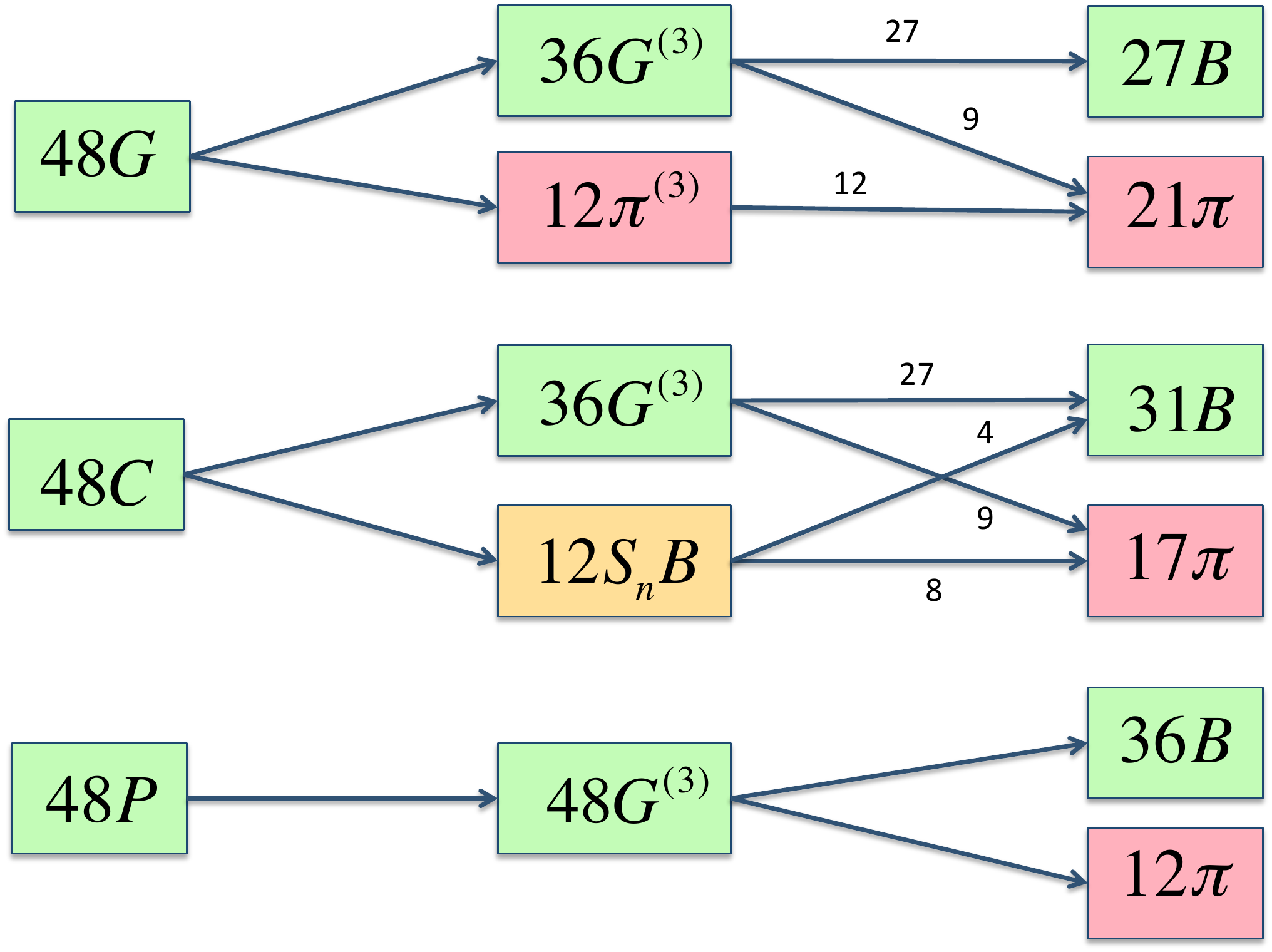}
  \caption{ Number of measurement paths per qutrit for each sequence of
  outcomes.}
  \label{Fig2}
\end{figure}

\subsection{Generalized Pauli Persistency}

Recall that the simple Pauli persistency is the minimum number of 
measurements, $\CN_{min}$, required to completely remove the
entanglement of the system.  One can define a generalized Pauli persistency 
as the number of measurements in an arbitrary sequence, $\CN(U_n,V_m)$, 
after which the entanglement is completely removed.  This number depends 
on the operators chosen for the first two measurements, but not on the third,
which removes any remaining entanglement.  In the case of the $C$ 
states, it also depends on the choice of the second qudit measured, given
the first (\ie, on $m-n$).

We define the average Pauli persistency over all measurement paths,
\be
   \CN_{ave} = \sum_{paths} \CN(U_n,V_m) \big( \sum_{paths} \big)^{-1}.
\label{PPave}
\ee
Since the choice of third measurement is irrelevant, as is the choice of
qudit for the first measurement, the average should be taken over a total
of $3(d+1)^2$ paths, 3 being the number of choices of the second qudit.
There are 48 such paths represented in Fig. 3 for qutrits.  The averages
over these paths are $\CN_{ave} = 2.31$, 2.65, and 2.75 for the $G$, 
$C$, and $P$ states, respectively.   As $d$ increases, these numbers all
increase and approach the value 3 asymptotically as $d \rightarrow \infty$. 
Although these numbers distinguish between the three states in the 
expected order, they are not dramatic.  The more interesting distinction is
in the difference between successes and failures in the final outcomes 
after two measurements.  With $d=3$, for example, the normalized 
difference, $\Delta = [\CN(B) - \CN(\pi)]/48$, takes the values 0.125,
0.292, and 0.50, for $G$, $C$, and $P$, respectively.

\subsection{Comment on Reduced States of Subsystems}

In Sec. III we derived the reduced density matrices of all subsystems
directly from the states.   In Appendix C we recover these results in operator
form (more concisely but with less insight), using the stabilizer formalism 
introduced here.   The main point of this is to show that a four-particle 
graph state is a 2-MM state if and only if none of its stabilizers has 
more that a single $I$ factor, a property unique to the $P$ states.  
\section{Conclusions and Open Questions}

We have shown that in four-particle systems, unlike two- and three-particle
systems, there is a qualitative jump in the potential for entanglement in going 
from binary to nonbinary cases.  This is because a third entanglement class is 
introduced for all prime $d \geq 3$ in addition to the two that exist for qubits.   
We have documented the quantitative differences among the three classes in 
terms of the reduced states of all subsystems, as well as the persistency of 
entanglement under measurement-induced steering.

The subsystem states (and particularly their purities as listed in Table I) show 
that $N=4$ is a transitional case, in the sense that there is a threshhold dimension 
($d=3$) for the existence of maximally multiparticle entangled states (MMES):  Only 
the $P$ states satisfy the requisite condition that all subsystems of up to half 
the system size are maximally mixed.  In fact, $N=4$ is the smallest such 
transitional number, and we expect that there exist larger systems ($N$) which are 
transitional in the same sense, whose threshhold dimensions could be larger than 
3.  Such transitions would open up new possibilities for maximal multiparticle
entanglement, a property known to be inaccessible to systems of 8 or more qubits 
\cite{Arnaud}.  It remains an unanswered question whether (i) there exist MMES
states of 7 or more particles of {\it any} $d$, and (ii) if so, then for some such $d$,
what are the limits on $N$ for their existence?

Both aspects of our study, but particularly the steering analysis, highlight the
extraordinary symmetries of the $P$ states with regard to permutations of both 
particles and measurement bases.   A consequence of these symmetries is that 
the three-particle GHZ states projected by first measurements on $P$ states
exhibit weak (or vulnerable) bases of all types (uniformly distributed over
$Z$ and $XZ^k$, $k = 0,1,...,d-1$), while those projected from $G$ and $C$ 
states are biased toward $Z$ or $X$ as weak bases.   Another general 
unanswered question is whether states with similarly high permutation 
symmetries will be found in higher-order transitional cases, exhibiting 
maximal entanglement at higher $N$ and $d$.

\appendix
\section{Graphs, states and stabilizers}
Here we review the formalism that relates graphs to states and stabilizers
through the so-called adjacency matrix, $\Gamma$, whose elements specify
the edge weights:  The edge connecting vertices $n$ and $m$ is assigned
the weight $\Gamma_{nm} = 0,1,...,d-1$ (where $0$ means no edge), so that 
diagonal elements are zero and $\Gamma_{nm} = \Gamma_{mn}$.   In our
examples, blue denotes the value 1, and red denotes $d-1= -1$.)  According
to the rules described in the text, the stabilizer generators are given by
\be
   g_n = X_n \bigotimes_m Z_m^{\Gamma_{nm}}, 
\label{Agenerators}
\ee  
and the stabilizers, labeled by the powers of the generators, are given by
\be
   S(p_1p_2p_3p_4) = g_1^{p_1}g_2^{p_2}g_3^{p_3}g_4^{p_4} = 
   \prod_n \big( X_n \bigotimes_m Z_m^{\Gamma_{nm}} \big)^{p_n}.
\label{Astabilizers1}
\ee
To re-express \ref{Astabilizers1} as an overall tensor product, we must reorder 
the factors to bring those operating on the same qudit together, 
\be
   S(p_1p_2p_3p_4) = \om^{\sum_{n>m} \Gamma_{nm} p_np_m}
   \bigotimes_{n=1}^4 X_n^{p_n} Z_n^{\sum_{m=1}^4 \Gamma_{nm} p_m},
\label{Atensorproduct}
\ee
where the phase prefactor results from the general commutators, 
$Z_m^b X_n^a = \om^{ab} X_n^a Z_m^b$.

The graph states are given in terms of standard basis states (here written 
$\ket{j_1j_2j_3j_4}$) by 
\be
  \ket{\psi} = \sum_{j_1j_2j_3j_4} \ket{j_1j_2j_3j_4} \exp \big( {2\pi i \over d}
 \sum_{n=1}^4  \Gamma_{nm}j_nj_m \big).
\label{Astate}
\ee
This is the joint eigenstate of all four generators with eigenvalues unity.  In
total, there are $d^4$ joint eigenstates with eigenvalues 
($e_1,e_2,e_3,e_4) \equiv (\om^{r_1},\om^{r_2},\om^{r_3},\om^{r_4}$), 
where $r_k = 0,1,...,d-1$, and we label their density matrices by these 
powers, $\rho^{(4)}(r_1r_2r_3r_4)$.   

The stabilizers (\ref{Astabilizers1}) have eigenvalues 
$\om^{r_1p_1+r_2p_2+r_3p_3+r_4p_4}$ in the states 
$\rho^{(4)}(r_1r_2r_3r_4)$, and so clearly their spectral representations are
\be
   S(p_1p_ 2p_3p_4) \equiv 
   \sum_{r_1r_2r_3r_4} \rho^{(4)}(r_1r_2r_3r_4) 
   \om^{r_1p_1+r_2p_2+r_3p_3+r_4p_4}. 
\label{Astabilizers}
\ee
This is a four dimensional Fourier transform on a hypercube with $d$ points on
a side.   Its inverse is
\be
   \rho^{(4)}(r_1r_2r_3r_4) = d^{-4} \sum_{p_1p_2p_3p_4} S(p_1p_ 2p_3p_4)
   \om^{-r_1p_1-r_2p_2-r_3p_3-r_4p_4}. 
\label{Aprojectors}
\ee
We immediately recover  \Eq{powers} of Sec. IV by writing S as the product of 
generators and recognizing the exponent as the product of $(e_n^*)^{p_n}$ 
factors. 
\bigskip

\section{Completeness of four-particle $G$, $C$, and $P$  
graphs for prime $d$}

Here we prove that an arbitrary four-particle graph is equivalent, under local
unitary transformations and permutations of qudits, to one of the $G$, $C$, 
or $P$ types shown in Fig. 1, for prime $d$.   Any such transformation on a 
graph state can be represented by combinations of two operations on its 
adjacency matrix $\Gamma$ \cite{BandB07}:   The first, $\circ_{n}(f)$, 
consists of multiplying the $n-$th row and $n-$th column by $f \neq 0$, 
\be
 \circ_n(f)~\Gamma_{lm} \equiv f^{\delta_{nl} + \delta_{nm}} \Gamma_{lm},
\label{circop}
\ee
while the second, $\ast_n(f)$, induces changes outside the $n-$th row 
and column,
\begin{equation}
    \ast_{n}(f)~\Gamma _{lm} = \Gamma _{lm} + f \Gamma _{ln}
     \Gamma_{nm}.
\label{astop}
\end{equation}
Two graphs are equivalent if their $\Gamma$ matrices are related by a 
sequence of these two operations together with permutations of qudits 
(interchanges of corresponding rows and columns).

Let us begin with the general six-edged graph shown in Fig. 3, in which all 
edge weights are nonvanishing.  The associated adjacency matrix is
\begin{equation}
  \Gamma = \left( 
\begin{array}{cccc}
0 & a & b & c \\ 
a & 0 & e & d \\ 
b & e & 0 & f \\ 
c & d & f & 0   
\end{array}%
\right).
\label{Gamma}
\end{equation}%
We can remove any element [say $d$, by the operation 
$\ast_3(-de^{-1}f^{-1})$] and scale two of the remaining elements to unity 
[say $\circ_3(e^{-1})$ then $\circ_4(ef^{-1})]$, leaving the five-edged graph, 
\begin{figure}
  \centering
    \includegraphics[scale=.7]{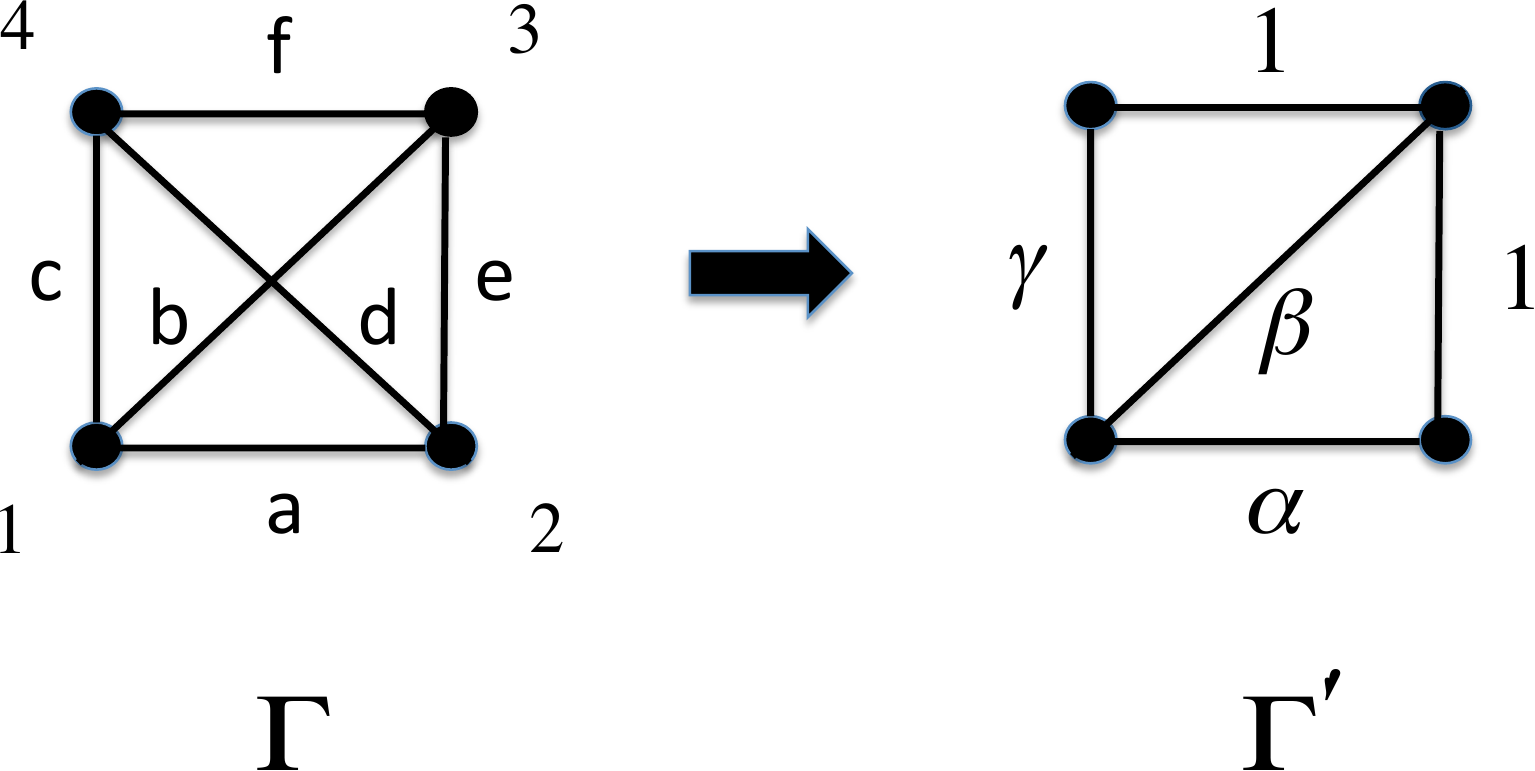}
  \caption{Most general six-edged graph and its reduction to an equivalant
  five-edged graph.}
  \label{Fig3}
\end{figure}
\begin{equation}
   \Gamma' = \left( 
\begin{array}{cccc}
0 & \al & \beta & \ga \\ 
\al & 0 & 1 & 0 \\ 
\beta & 1 & 0 & 1 \\ 
\ga & 0 & 1 & 0%
\end{array}%
\right) ,
\label{Gammap}
\end{equation}%
where the operations above dictate that
\bea
  &  \al = a - bdf^{-1},                 \cr
  &  \beta = be^{-1},                   \cr
  &  \ga = (ce - bd)f^{-1}.
\label{parameters}
\eea
Note that either or both of $\al$ and $\ga$ may vanish without any of the original
parameters vanishing.   If both vanish, \underline{$\al = \ga = 0$}, then the 
permutation ($3 \leftrightarrow 4$) and scaling $\circ_1(\beta^{-1})$ reduce
$\Gamma'$ to
\begin{equation}
   \Gamma_G = \left( 
\begin{array}{cccc}
0 & 0 & 0 & 1 \\ 
0 & 0 & 0 & 1 \\ 
0 & 0 & 0 & 1 \\ 
1 & 1 & 1 & 0%
\end{array}%
\right) ,   
\label{GammaG}
\end{equation}%
\newline
which is equivalent to the GHZ graph in Fig. 1.   Now suppose that at least one 
of $\al$ and $\ga$ is nonzero.  Since they are interchangeable (by 
$2 \leftrightarrow 4$) choose \underline{$\al \neq 0$}:  Then the operation 
$\ast_2(-\beta \al^{-1})$ removes $\beta$, and the scaling $\circ_1(\al^{-1})$ 
leaves
\begin{equation}
   \Gamma(\tga) = \left( 
\begin{array}{cccc}
0 & 1 & 0 & \tilde{\ga} \\ 
1 & 0 & 1 & 0 \\ 
0 & 1 & 0 & 1 \\ 
\tilde{\ga} & 0 & 1 & 0%
\end{array}%
\right) ,
\label{GammaCP}
\end{equation}%
\newline
where
\be
   \tilde{\ga} = \ga/\al = (ce-bd)/(af-bd), 
\label{gammatilde}
\ee
which is well-defined because $\al \neq 0$.  Clearly $\tga = 1$ corresponds to the
$C$ graph of Fig. 1. 

We can show immediately that the three-sided graph associated with $\tga = 0$ 
is equivalent to this:  Its state is given by \Eq{Ca} with the phase factor changed to 
$\om^{j(i+k) + lk}$.  The $j$ and $l$ sums can be identified as Fourier transforms 
as before, but now the resulting state is
\be
   |C''\rangle = {1 \over d} \sum_{i,k}|i,i+k,k,k\rangle.  
\label{Cc}
\ee
Comparing with the $\ket{C'}$ expression (\ref{Cb}), both have repeated indices, 
but in different places.  So the permutation ($2 \leftrightarrow 3$) followed by simple 
variable changes maps \Eq{Cc} into \ref{Cb}.  

All remaining cases correspond to the remaining values of $\tga$, namely 
$2,...,d-1$.  One can see immediately that the $\tga = -1$ case, with permutations
$1 \leftrightarrow 2$ and $3 \leftrightarrow 4$,  corresponds to the $P$ state defined 
by Fig. 1.   Appendix C shows formally that all of the values, $\tga = 2, ...,d-1$, have 
the same entanglement characteristics, by calculating the reduced density matrices 
as functions of $\tga$.  One reaches the same conclusion with the approach of 
Sec. III, using the general form (\Eq{psib}) for $C$ and $P$ states (repeated here):
\be
    |\psi(\tga) \rangle = {1 \over d} \sum_{i,k}|i,i+\tga k,k,i+k\rangle.  
\label{Pc}
\ee
For maximal mixing of all pairs (the 2-MM property) we require that the basis states
in $A$ span $\CH_A$, and that their associated states span $\CH_B$, for any 
choice of the pair, $A = n,m$.  This can happen if and only if $\tga = 2,...,d-1$ and 
$d$ is a prime.  The other values, $\tga = 0$ and 1, allow repeated indices which 
cause one of the pairings to fail.

We have shown thus far that an arbitrary six-edged graph must reduce to one of the 
$G$, $C$, or $P$ classes.  This proof incorporates some, but not all, lower-edged 
graphs.  Here we sketch proofs that cover them all.

\noindent (i) Five-edged graphs:  All six placements of the 0 element are equivalent
under permutations $n \leftrightarrow m$, so choose $\Gamma_{24} \equiv d=0$ (in the 
notation of \ref{Gamma}).  Then remove $b$ using $\ast_2(-b/ae)$ and scale $a$, $e$,
and $f$ to unity.  This leaves $\Gamma (\tga)$ in the form of \Eq{GammaCP}, with 
$\tga = ce/af$ nonzero, which is equivalent to a $C$ or $P$ state depending on the 
value of $\tga$.  (Note that no five-edged graph can reduce to a GHZ state.)

\noindent (ii) Four-edged graphs:  There are 15 ways to place two zeroes in the matrix.  
In three of these, the zeroes are diagonally-coordinated.  These three are equivalent 
under permutations alone, and one of them is $\Gamma(\tga)$ with $\tga$ nonzero.  
The remaining 12 graphs are identical among themselves under permutations, so 
choose $c = d = 0$.   Again remove $b$ using $\ast_2(-b/ae)$ and scale the 
remaining three elements to unity, leaving $\Gamma(\tga)$, now with $\tga = 0$, 
a $C$ state.  (Note that no four-edged graph reduces to a GHZ state.)

\noindent (iii) Three-edged graphs:  Since there are 20 ways to place three zeroes in 
the matrix, let us focus instead on the graphs themselves.  Clearly there are four GHZ 
graphs in which a single vertex connects to three edges.   These four are related by 
cyclic permutations, or rotations of Fig. 1a.   It is easy to see (without enumerating) 
that all remaining graphs must have two vertices each connected to two edges, and 
two vertices each connected to single edges.  The vertices can be rearranged and 
aligned so that the edges form a straight line without overlapping.   Starting with this 
line, one may count 16 ways of placing the vertices back on a square (making four 
distinct shapes, each with four orientations).  All 16 are permutation-equivalents 
to $\Gamma(\tga)$ with $\tga = 0$, a $C$ state.
%
%As a final note, one can see from the above conditions on $\tga$ that when $d$ 
%is composite, the proof of the completeness of $G$, $C$, and $P$ classes fails.  
%We shall describe the consequences in Appendix D.
%
%\section{Composite dimensions and entanglement}
%
%Composite-$d$ systems have the same $G$, $C$, and $P$ states as written in 
%Sec. II,  and the $G$ and $C$ states have the same entanglement properties
%discussed in Sec. III.  However, the composite-$d$ $P$ states may no longer 
%be MMES.  In particular, if one adopts the more general form of (\Eq{Pc}) above, 
%then with $\tga$-values outside the range (1, 0), the entanglement properties 
%depend on the choice of $\tga$.  Below we identify these classes [*[maybe]*]
%and show that their entanglement measures lie between those of $C$ and $P$ 
%states.
%

\section{Reduced density matrices - stabilizer expansions}

Here we rederive the reduced density matrices of Sec III from the stabilizer 
formalism and obtain the results in purely operator form.   We begin with the pure
four-particle density matrix as given by \Eqs{Atensorproduct}{Aprojectors}, with 
eigenvalues taken as unity, 
\be
   \rho^{(4)} = d^{-4} \sum_{p_1p_2p_3p_4} \om^{\sum_{n>m} \Gamma_{nm} p_np_m}
   \bigotimes_{n=1}^4 X_n^{p_n}
   Z_n^{\sum_{m=1}^4  \Gamma_{nm} p_m}.
\label{stateC}
\ee
Specializing first to the $G$ state with $\Gamma_G$ given by \Eq{GammaG},
this density matrix (letting $p_1$...$p_4 \rightarrow a,b,c,d$) becomes
\be
   \rho(G) = d^{-4} \sum_{abcd} \om^{(a+b+c)d} (X^aZ^d)\otimes(X^bZ^d)
   \otimes(X^cZ^d)\otimes(X^dZ^{a+b+c}).
\label{GHZC}
\ee
Tracing over any particle would give an equivalent result, so choose $n=4$; 
surviving terms have $d=0$ and $a+b+c=0$, so that
\be
  \hbox{Tr}_4 [\rho(G)] = d^{-3} \sum_{ab} X^a \otimes X^b 
  \otimes X^{-a-b}.
\label{GHZC1}
\ee
The choice of second particle is again immaterial, so choose $n=3$; the result is
\be
  \rho_{1,2} =  \hbox{Tr}_{3,4} [\rho(G)] = d^{-2} \sum_a X^a \otimes X^{-a} .
\label{GHZC2}
\ee
This shows that no two-particle subsystem is maximally mixed in the $G$ state.  

We now treat the $C$ and $P$ states simultaneously using $\Gamma(\tga)$ of
\Eq{GammaCP}.   The density matrix (henceforth replacing $\tga \rightarrow \ga$) is
\be
   \rho(\ga) = d^{-4} \sum_{abcd} \om^{ab+bc+cd+\ga ad} (X^aZ^{b+\ga d}) 
   \otimes (X^bZ^{a+c}) \otimes (X^cZ^{b+d}) \otimes (X^dZ^{\ga a+c}).
\label{PC}
\ee  
Tracing over the fourth particle removes all but the $d=0$ and $c= -\ga a$ terms, 
resulting in
\be
   \hbox{Tr}_4 [\rho(\ga)] = d^{-3} \sum_{ab} \om^{(1-\ga)ab} (X^a Z^b) 
   \otimes (X^bZ^{(1-\ga)a}) \otimes (X^{-\ga a} Z^b).
\label{PC1}
\ee
The choice of second particle now matters, and we must treat three(!) cases separately:

\noindent (i) Tracing over particle 1 (the simplest case, with $a=b=0$) gives us 
\be
   \rho_{2,3} = \hbox{Tr}_{1,4} [\rho(\ga)] = d^{-2} I \otimes I,
\label{PC2I}
\ee
showing that the subsystem (2,3) is maximally mixed.   

\noindent (ii) Tracing instead over particle 2, we get
\be
  \rho_{1,3} = \hbox{Tr}_{2,4} [\rho(\ga)] = d^{-2} \sum_{d} X^a
  \otimes X^{-\ga a}  \delta_{(1-\ga) a,0}.
\label{PC2ii}
\ee
This reduces to $d^{-2}I \otimes I$ unless $\ga = 1$ (the four-sided $C$ state).

\noindent (iii) Tracing finally over particle 3 gives us 
\be
  \rho_{1,2} = \hbox{Tr}_{3,4} [\rho(\ga)] = d^{-2} \sum_{d} X^a
  \otimes Z^{(1-\ga)a} \delta_{\ga a,0}, 
\label{PC2iii}
\ee
which also reduces to $d^{-2}I \otimes I$, but now unless $\ga = 0$ (the three-sided 
$C$ state).  Cases (i-iii) taken together show that all two-particle subsystems are 
maximally mixed for the $P$-states, $\ga = 2,...,d-1$, while for $\ga = 0$ or 
1 ($C$ states), only two of the three pairs are maximally mixed.   

These outcomes may be traced to the fact that no $P$ stabilizer has more than a 
single $I$ factor (proved immediately below), while some $C$ stabilizers do:  For 
$\ga = 1$, the $I$ factors are diagonally opposed (case ii), while for $\ga = 0$ they 
appear together, on qudits 1,2 in one instance and on 3,4 in the other (case iii).  
In the case of $G$ states, there are stabilizers in which $I$ factors appear on an 
arbitrary pair of qudits, so that no subsystem of two particles is maximally mixed.

\subsection{Proof that $P$ stabilizers have at most single $I$ factors}

The generators dictated by $\Gamma(\tga)$ of \Eq{GammaCP} 
(again letting $\tga \rightarrow \ga$) are
\begin{equation*}
     XZIZ^{\ga}, \hskip0.7truecm  ZXZI,  \hskip0.7truecm  IZXZ,
     \hskip0.7truecm   Z^{\ga}IZX.
\label{genB}
\end{equation*}%
The resulting stabilizers (ignoring the commutation-generated phase factors) are
\begin{equation}
   S(a,b,c,d) = (X^aZ^{b + \ga d}) \otimes (X^bZ^{a + c}) \otimes 
   (X^cZ^{b+d}) \otimes (X^dZ^{\ga a+c}).
\label{stabB}
\end{equation}%
For comparison, let us first identify $C$-state stabilizers with more than
single $I$ factors:

\noindent (i) If $\ga=0$, then $g_1 = S(1,0,0,0)$ and $g_4 = S(0,0,0,1)$ 
each have two $I$ factors.  

\noindent (ii) If $\ga=1$, then $g_1g_3^{-1} = S(1,0,-1,0)$ and 
$g_2g_4^{-1} = S(0,1,0,-1)$ each have two $I$ factors.

\noindent Now for the $P$ states:

\noindent (iii) If $\ga$ takes any other value (2,..., $d-1$), then one can write out  
each of six possible conditions for the existence of at least two $I$ factors, and
show immediately that none of these has a nontrivial solution.
%not find a stabilizer with more than a single $\CI$ factor, because neither 
%condition for the existence for two such factors, 
%\{$a=0$, $c=0$, $b+d=0$, $d+b\Gamma =0$\} nor
%\{$b=0$, $d=0$, $a+c=0$, $c+a\Gamma =0$\}, has a nontrivial solution.

This proves that no $P$-state stabilizer has more than a single $I$ factor.  This 
property is a necessary and sufficient condition for $P$ states to have the 2-MM 
property, because then and only then, tracing over the states of any two particles
annihilates all of the stabilizers, so that only the identity $I \otimes I$ survives.  


\begin{thebibliography}{99}
\bibitem{observable}  We refer to $U$ and other Pauli factors as observables,
with the understanding that, being unitary, they are exponentials of hermitian 
operators which have real eigenvalues, $a = 0,1,...,d-1$, the eigenvalues of 
$U$ being $\om^a$.
\bibitem{GHZ3}  One can see that this generator set is related by Fourier
transforms to that defined by the three-qudit GHZ graph.  Alternatively, one may 
verify immediately that the three-particle analog of \Eq{GHZb} is an eigenstate 
of this generator set.
\bibitem{nonseparable} One can easily show that a separable three-particle
eigenstate of Pauli operators must have stabilizers with two $I$ factors \cite{JL2}.
Nontrivial examples are found in the case of C-states, to follow.
\end{thebibliography}
\end{document}